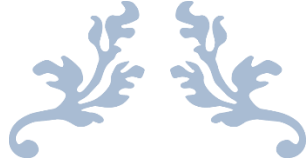

# IDENTITY PROVE LIMITED INFROMATION GOVERNANCE POLICY AGAINST CYBER SECURITY PERSISTENT THREATS

## Antigoni Kruti

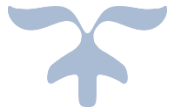

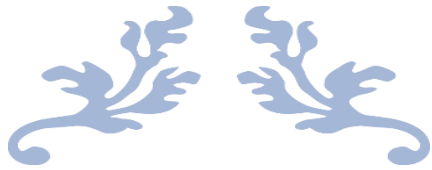

# INFORMATION GOVERNANCE POLICY

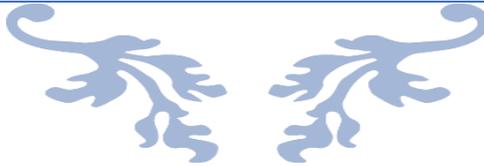

## 1. Introduction

**1.1** Identity Prove Limited (IDPL) is a long-founded online identity verification software provider of citizens for Banking services. IDPL applies an information governance based on the ISO/IEC 27001:2022 standard of security and within GDPR to accomplish face verification.

**1.2** The company has a good reputation for biometric authentication services that allow a secure, simple, sustainable online access for financial services providers on delivering security device-independent, ensuring reassurance and convenience to users. The company should ensure a right person, a real person, authenticating in real-time.

**1.3** The IDPL company must assume sustainable security models for the duration of day-to-day operations does not involve human intervention. The IDPL's Security Operations Centre (ISOC) should continuously provide the optimum scale of system performance, utilize security procedures against new threats, ensure the optimum scale of system performance capabilities.

## 2. Purpose

**2.1** The aim of information governance policy is to declare and to demonstrate the performance of the company on effectively and efficiently way in front of risk detection and vulnerability mitigation.

## 3. Scope

**3.1** The scope of this policy involves all management systems and stakeholder's details, include unique identifiers of submitter and receiver.

**3.2** The company has in-house systems focused on all potential risks to client data and its information system assets.

## 4. The Policy Objectives

**4.1** The company regularly monitor the deployment system and the environment to utilize that the company cybersecurity status is maintained.

i. Information Governance policy aligns the cyber security services with IDPL target requirements that are related with confidentiality, availability, and integrity of system recorded data.

ii. Prioritize resources, examine staff activities, and execute problem resolutions in accordance with established processes.

iii. Proceed lawfully a transparent mechanism for the data subject.

iv. Minimize adequate, limited, and relevant data in relation to the purpose.

v. Process data including protection against unlawful or unauthorized destruction, damage or accidental lost.

vi. Observe the deployment environment of the systems on cyber security management and maintain status.

vii. Forbid unauthorized operation and train staff to be confidential and vigilant on data security.

viii. Incorporate security rules agreements into business associate contracts mandatory by the privacy rules.

4.2 IDPL should provide the necessary resources and document plans to detect and mitigate the risks on the company.

## 5. Policy Framework

5.1 Culture IDPL develops a first security culture between stakeholders for all activities include security as a priority.

5.2 IG team must be cross trained to develop effective communication with stakeholders and must be able to cooperate in several viewpoints through Role-Based Access Policy.

5.3 Security Awareness Training (SAT) must consistently educate the employee for potential schemes to prevent cyber security attacks. The employee should not leave the IDPL vulnerable to data breach scenarios.

5.4 The company governs and comply data on accordance with Digital Access Requirements

(DSAR) under GDPR requirements.

## 6. Information Governance policies and procedures

6.1 This information governance policy is developed through comprehensive examinations of policies and procedures involving:

   i.   Contactless online identity verification procedure

   ii.  Electronic Identification, Authentication and Trust Services (EIDAS) procedure

   iii. Long-term digital preservation procedure

   iv.  Costumer Diligence procedure

   v.   Know Your Costumer policy

   vi.  Anti-Money Loundering policy

   vii. Bring Your Own Device policy

   viii. Equipment disposal and asset management policy.

## 7. IDPL monitoring measures and review mechanisms

7.1 The company must annually examine all policy components sustainability in accordance with threats and risks.

7.2 IDPL company should give the employee the opportunity to satisfy their tastes by allowing the usage of personal devices by applying Bring Your Own Device (BYOD) policy.

## 8. Compliance

8.1 Staff must adhere all the policy and procedure of company in accordance with discipline regulation.

## 9. Approval

9.1 On the title page of this policy is shown the revision number and the date of approval by the bord.

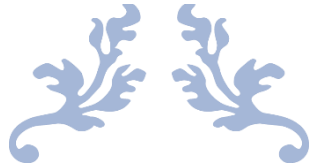

# INFORMATION GOVERNANCE POLICY REPORT

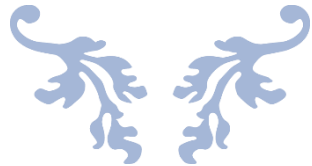

| Version | 1.0 |
| --- | --- |
| Author | Antigoni Kruti |
| Date issued | 25 July 2022 |
| Revied date | 3 September 2023 |
| Target audience | Identity Prove Limited company staff and stakeholders |
| Word counts | 3954 |

Table of content

## 1. Identity Prove Limited Background and Information Governance Policy context

Identity Prove Limited (IDPL) is an identity service provider company. It started to operate for the first time on 17 August 2020 in London, United Kingdom. The primary function of the company is to optimize the digital ID verification on Banking services through Inclusion and Application Standards for Automated Facial Analysis Technology, IEEEP70013 in compliant with the ISO/IEC 27001:2022 standard of security within General Data Protection Regulation to achieve face verification supported by revolutionary multimodal ID-platform.

The company's vision is to comprise a robust facial recognition system that fraud prevention is cost-effectively accessible to every business and ID verification can be accomplished in the whole countries in milliseconds. The mission of accomplishment biometric verification and authentication consist of liveness perception by capturing minor facial features, 3D depth detection to capture live biometric data for accurate matching, synthetic identity fraud prohibitions using deep algorithm, and fast authentication and mapping techniques powered by AI.

## 2. Executive Summary
### 2.1 Introduction

This report is academic research to estimate a critical understanding that analyses and examines the necessary IDPL Information Governance Policy to monitor staff legality by improving transparency of data usage, to make accountable employees personal and financial data by processing data subcontractors, or controllers, and to progress cooperation between data protection establishments and IDPL company by adopting sanctions and taking joint decisions. IDP has made a huge effort to provide all engagements of GDPR legal counsel on compliance with proper liabilities to ensure usability, privacy, security, and scalability.

### 2.2 Report structure

The academic report begins with an introduction of the main functions and the background of IDPL company and its Information Governance policy. Then it gives critical examinations of implementing every specific policy and analysing each component into the company.

## 2.3 Why should Identity Prove Limited implement an IG Policy

IDPL must implement an effective Information Governance Policy to establish an organizational culture with an internal effective combination of staff ethical and legal attitudes and external conduction of response activities to stakeholders. As a result of successfully IG policy implementation, the company has prevented risk tolerance of the cyber breach and has dictated cyber breach response program. Furthermore, IG policy refers operational day to day development such as the adjustment of standard operating procedures (SOPs) combined with key performance indicators (KPIs) to automate repeatable tasks that involves stakeholders.

## 3. Benefits of Information Governance implementation

"Information governance is a sort of discipline that has emerged because of new and tightened legislation governing businesses, privacy concerns, legal demands, external pressures such as hacking and data breaches, and the recognition that multiple overlapping disciplines (Blum, 2020)." Information Governance key concepts are data privacy, information security, regulatory compliance, data governance, long term digital preservation (LTDP) and record information management (RIM). The accomplishment of those concepts is related with advanced technology through disaster recovery (DR), knowledge management and business continuity (BR).

Information Governance Policy is vital for IDPL company to proactively manage e-discovery information reduction and to simply the volume of task responsive information. IG policy helps employees to separate knowledge by focusing on the most valuable information through the development of productivity for all company structure components. In this way, it helps the company to reduce the complexity and the costs of IT environment by getting rid of unnecessary data or information with defensible mechanisms, that reduces the risks of data management failures. Furthermore, IG policy legislation assists on innovative manners that asset on security and privacy compliance stakeholder improvements.

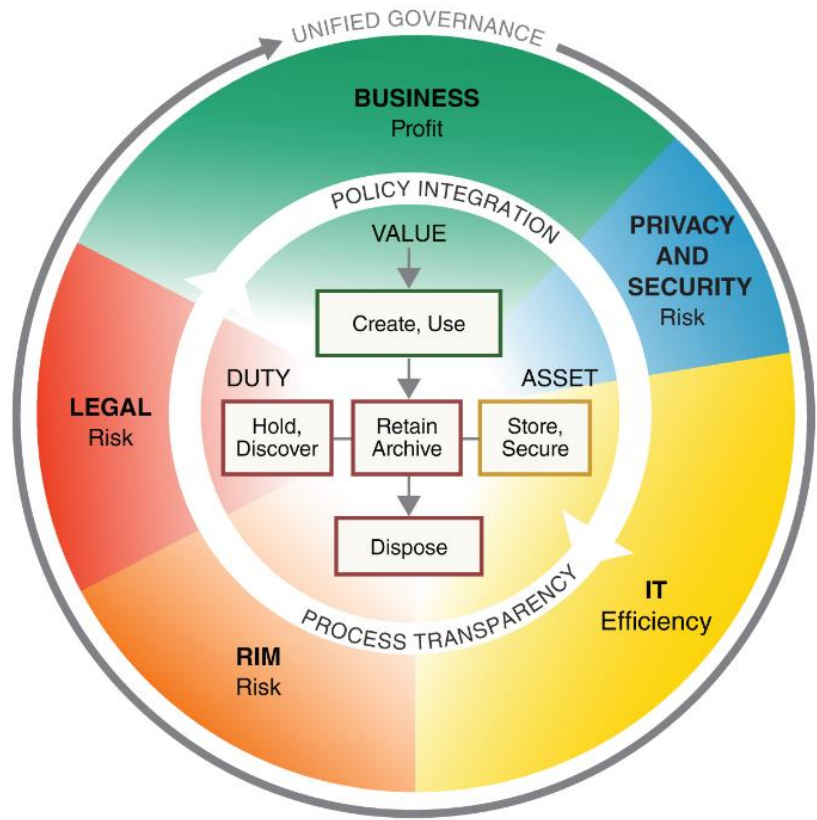

*Figure 1 The Information Governance Policy model (Roberts, A., 2021)*

The Information Governance Policy of the IDPL company has implemented GDPR to achieve the strength of individual rights on data usage transparency and personal data portability. It develops high protection between data authorities which involves joint decision adoptions of processing operational and transactional actions on personal and financial data that include GNI, financial transactions and human records (Wolfe, 2020). IDPL company has a compliance plan that includes organisation for data processing, collection, and security access management. The compliance of GDR requires deep understanding of data operations, IT procedures and systems related to impact analysis of data protection, user explication consent development, advanced data security technologies. On the regulation body documentation of any security failure must be detected within 72 hours to establish knowledge regard to taking optimize solutions of problems.

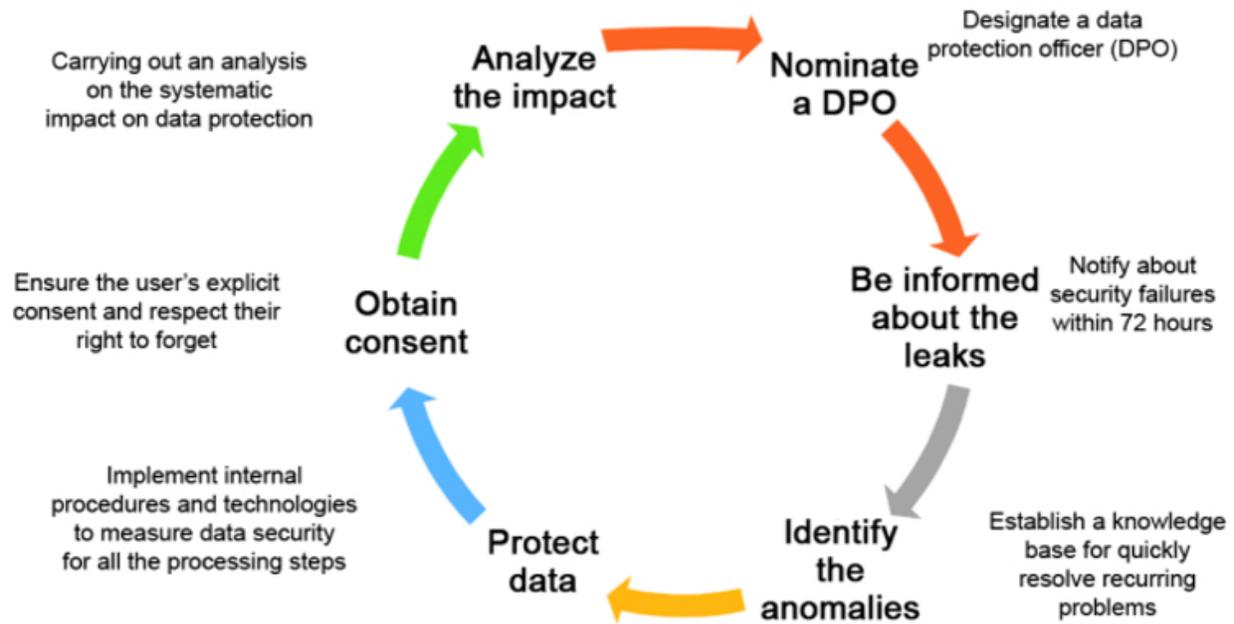

*Figure 2 The GDP model (Smallwood, R.F., 2019)*

4. Policy elements analysis

Policy 1.1 - "Identity Prove Limited (IDPL) is a long-founded online identity verification software provider of citizens to the Banking services. IDPL applies an information governance based on ISO/IEC 27001:2022 standard of security and within GDPR to accomplish face verification."

Analysis of 1.1 - IDPL customers are banking service companies, and it provides legalized services by the Jersey Financial Service Committee. IDPL must understand the requirements and meet the standards of clients to promote liveness detection, anti-spoofing checks, fake image detection and 3D depth identity perception. The company has succeeded to develop 5 seconds verification mechanism for costumers, to detect spoof attacks and to create a strong defence against stolen identities and synthetic identity frauds. The company has used the Application Standards for Automated Facial Analysis Technology (IEEEP70013) to implement the Facial Action Coding System (FACS) from software pipeline. FACS utilizes gradient-based descriptors, binarized local texture, two-layers appearance descriptors and patio-temporal appearance features, implemented by AdaBoost algorithm (Blum, 2020). It guarantees face tracking and detecting, image feature extraction and classification.

Policy 1.2 - "The company has a good reputation for biometric authentication services that allow secure, simple, sustainable online access for financial services providers on delivering security device-independent, ensuring reassurance and convenience to users. The company should ensure right person, a real person, authenticating in real-time."

Analysis of 1.2 - IDPL ensures cryptography, behavioural science, advanced machine learning and optics unique digital Biometric Authentication products. The automated method of digital ID cards verification extracts the information incorporated in files through Optical Character Recognition technology (OCR) to lead costumers onboarding with IDPL services. IDPL must create a completely contactless verification process, integrated with AI-powered document authentication, adopted on mobile verification Android SDK or IOS integration (Roberts, 2021). E-ID cards have Nar Field Communications (NFC) chip that contains the digitally signed and encrypted data, which make them highly impenetrable and secured. IDPL assists in identity verification solutions that provide ultra-reliable security for business and end-users. Then, the system starts to scan the Machine-Readable Zone (MRZ) incorporated on documents. Authenticity is checked on the MRZ section and if the NFC chip is matched, its data is extracted and shown to the end-user device screen. The primary authentication mechanism that IDPL must use is "Something you are", it comes with biometric authentication such as iris or facial recognition, fingerprint reading detection. Another method of authentication used by IDPL company is two factor-authentication (2FA), which is particularly useful for creating accounts and resetting user passwords. During the authentication process the user receives a one-time passcode (OTP) via an application or SMS. If a device is hacked, the attacker would not be able to have access in a certain application if it is protected with another layer of authentication. The physical features of the costumers must be used for biometric authentication procedures.

Policy 1.3 - "The IDPL company must assume sustainable security models for the duration of day-to-day operations does not involve human intervention. The IDPL's Security Operations Centre (ISOC) should continuously provide the optimum scale of system performance, utilize security procedures against new threats, ensure the optimum scale of system performance capabilities."

Analysis of 1.3 - ISOC is responsible for cyber breach response operating activities include scientific inspection of compromised systems, data acquisition and preservation. Furthermore, continuously monitoring and reporting are the crucial components of operations to enhance the performance of breach response facilities. IDPL consider essential the ISOC response program and it consists of reporting security events and weaknesses, maintaining responsibilities and procedures, decision making on security issues, responding to incidents, learning from incidents.

Business stakeholders and cyber security experts have different perspectives on cyber security and various technology to communicate with each other. IDPL must implement a target operating model (TOM) to ensure that cyber security resources associate with business targets. TOM helps to establish a sufficient deployment of resources, to design a cyber security model focuses on risks, to analyse and improve continually the organization structure of each competency. All processing of biometric data occurs on cloud-based servers not on costumers' devices so devices cannot be compromised. IDPL uses Privacy Firewall to retain and control all Personally Identifiable Information (PII). The company priority is to ensure sustainability on costumers' privacy by operating with the highest privacy standards.

## 2. Purpose

Policy 2.1 - "The aim of information governance policy is to declare and to demonstrate the performance of company on effectively and efficiently way in front of risk detection and vulnerability mitigation."

   I.   use individual data to prove citizens identity,

  II.   protect costumers against online and offline threats,

 III.   maximize the contactless online identity verification,

  IV.   reduce time and cost of administration,

   V.   increase inclusivity and accessibility, scalability,

  VI.   deliver security, usability, and privacy."

Analysis of 2.1 - The policy define the role and main functions of IDPL.

## 3. Scope

Policy 3.1 - "The scope of this policy involves all management systems and stakeholder's details, include unique identifiers of submitter and receiver."

Analysis of 3.1 - This scope states Information Governance policy implementation on company activities involve with stakeholders

i. Law enforcement agencies: National Security Agencies, Policy, Military, Internal Quality Assurance Mechanisms, and government agencies.
ii. Legal Representations: Internal Quality Assurance Mechanisms, Prosecutors, Defence
iii. Audit Agencies: Legal Assurance Agencies

Policy 3.2 - "The company has in-house systems focused on all potential risks to client data and its information assets systems."

Analysis of 3.2- IDPL bring its cyber-security in house system developed by Security Operation Centre (SOC). It assumes staff training from the company specific needs, reporting on a regular basis to embed a safe cyber security structure and creating a working culture for company. In house pros involves designing the security procedures and monitoring capacities that best meet the organisation's requirements. The company purchased several AI-driven security tools, so CISOs and security employees plan and implement significant investments on company. IDPL track abilities, which are deposited on-site and collected event log data. The external data transfer is expected to be reported for safety measures. IDPL ensure breach transparencies and coordinate event reaction process in-house and develops an integrated security strategy.

4. The Policy Objectives

Policy 4.1 - "The company regularly monitor the deployment system and the environment to utilize that the company cybersecurity status is maintained."

ix. Information Governance policy aligns the cyber security services with IDPL target requirements that are related with confidentiality, availability, and integrity of the system recorded data.

x. Prioritize resources, examine staff activities, and execute problem resolutions in accordance with established processes.

xi. Proceed lawfully a transparent mechanism for the data subject.

xii. Minimize adequate, limited, and relevant data in relation to the purpose.

xiii. Process data including protection against unlawful or unauthorized destruction, damage or accidental lost.

xiv. Observe the deployment environment of the systems on cyber security management and maintain status.

xv. Forbid unauthorized operation and train staff to be confidential and vigilant on data security.

xvi. Incorporate security rules agreements into business associate contracts mandatory by the privacy rules."

Analysis of 4.1 - IDPL Information Governance policy identifies and evaluates the resilience of the system availability, confidentiality, integrity to detect and mitigate weakness, risks, threatens for customers by developing sustainable plans of the company system.

Policy 4.2 - "IDPL should provide the necessary resources and document plans to detect and mitigate the risks on the company."

Analysis of 4.2 - IDPL has established the Cyber Incident Response Plan, which defines the way that cyber incident is handed through reporting, analysing, and mitigating the security incident. The incident must be reported within 1 hour to IT Service Request Desk, which is as a central point of contact to report any suspicious action insides the security breach of personal information (Smallwood, 2020). Then, the reported documentation will be sent as Priority Incident Request (PIR) to Chief Information Security Officer. CISO preliminary analyse the facts and assesses the incident nature and scope. Also, CISO provides an overview of the current situation for Security Manger by identifying the system components affected and determining the suspected breach of personal data. Then if the suspected breach is confirmed, the incident request will be updated from Service Request Desk Incident Response Team Activation – Critical Security Problem and Fist-Level Escalation Members associate with Disaster Recover (DR) procedures. They will determine, what, where and how the incident breach occurred by considering third party connections, systems, and audit logs components to take measures of controlling and preventing unauthorized rogue actions.

5. Policy Framework

Policy 5.1 - "Culture IDPL develops a firs security culture between stakeholders for all activities include security as a priority."

Analysis of 5.1 - Culture impacts the way that IDPL breach response and the involvement of stakeholders in response cooperative activities within ISO27001 security standard. The company supports its Observational Scaled Assessment of Teamwork (OAT) to asset leadership and coordinate security challenges within cyber defence team. IDPL measures strong leadership, communication, and collaboration between stakeholders.

Policy 5.2 - "IG team must be cross trained to develop effective communication with stakeholders and must be able to cooperate in several viewpoints through Role-Based Access Policy."

Analysis of 5.2 - According to Role-Based Access Policy the board of IDPL has decided that all staff members are responsible for their own areas and duties that they are involved in the project (RBAC). According to Gorecki, 2020 the role of Chief Information Governance Officer (CIGO) is to lead security team members toward task accomplishments and implementations of key IG policy features that are authorized by executive stakeholders and are assigned appropriately to employees' functional expertise area. CIGO responsibility is to determine legal requirement for day-to-day IG program. Data Protection Officer (DPO) preserves the connection between IDPL company and European authorities. DPO is responsible for data processing decisions and activities. The company is particularly careful about the contact, DPO must be able to accomplish all the needs of the organisations. Senior Record Manager (SRM) is responsible to lead, plan and manage the department record management program. SRM directs staff to implement the record management program through proper training related to principles and requirements of data security.

Policy 5.3 - "Security Awareness Training (SAT) must consistently educate the employee for potential schemes to prevent cyber security attacks. Employee should not leave the IDPL vulnerable to data breach scenarios."

Analysis of 5.3 - IDPL invests in security monitoring, risks in education enforcement and Security Awareness Training (SAT) through IG policy to decrease IP leaks or to prevent the malicious user. SAT increases effective engagement and user awareness with the help of the worldwide information security standard, ISO/IEC 27001. IDPL implements Document Life Cycle Security (DLS) technology such as Information Rights Managements (IRM) to prevent malicious users or threat vectors on information assets. IDPL must execute SAT as an ongoing process because it is not a one-and-done activity (Fréminville, 2020). IG policy is on accordance to meet OSHA requirements that engage employees with new opportunities and integrate new content on a consistent basis product, which is beyond all traditional computer-based activities. SAT involves role-based trainings for optimal customization on the company environment with interactive content, gamification methods to keep learners engaged on the robust library and flexible micro learning data security issues, integration with end-point systems combined with metrics to monitor employee participation and performance.

Policy 5.4 - "The company govern and comply data on accordance with Digital Access Requirements (DSAR) under GDPR requirements."

Analysis of 5.4 - Information Life Cycle Management (ILM) optimally and appropriately manages the information in different stages: creation, meeting legal requirements and distribution. Another crucial process for successful IG is Master Data Management (MDM), it deals with data integrity issues such as data accuracy. IDP uses Data Loss Prevention (DLP) and data mapping to detect the flow of information that should be monitored and analysed with security technology. IRM protection is added from the company to provide an audit trail of documents scalability scale, to integrate enterprise-wide system, for instance ECM, e-discovery, Product Life Cycle Management, and planning of resources. One example of IRM that IDPL uses is Payment Card Industry Data Security Standards (PCI-DSS), which is the strict regulation intended for processors and credit cards (Leszczyna, 2018). It comes from highly regulated and secured entities.

6.Information Governance policies and procedures

Policy 6.1 - "This information governance policy is developed through comprehensive examinations of policies and procedures involving:

ix. Contactless online identity verification procedures

x. Electronic Identification, Authentication and Trust Services (EIDAS) procedures

xi. Long-term digital preservation procedures

xii. Costumer Diligence procedure

xiii. Know Your Costumer policy

xiv. Anti-Money Loundering policy

xv. Personnel security procedures

xvi. Bring Your Own Device policy

xvii. Equipment disposal and asset management policy."

Analysis of 6.1 - IDPL has KYC policy to achieve the identification of the costumer, verification of identity, monitoring all the regular activities of customers and analyse the source fundings and activities. This policy applies Costumer Due Diligence (CDD) procedure, which is the control procedure to prevent threads and risks of new and existing customers. IDPL uses CDD to asset and monitor and examine the risk profile of costumers into occasional transactions or business relationship. Hence, Anti-Money Laundering policy presents a set of regulations and rules that hinder criminals from laundering. It involves procedures and laws to detect and identify Counter Financing Terrorism (CFT) supported by Financial Action Task Force (FATF) purpose to ensure globally comprehensive international standards (Roberts, 2021). IDPL has chosen the AML implementation to monitor and control the risk level and perception of its customers through Suspicious Activity Report (SAR). IDPL company uses SAR to check tax evasion, terrorist financing and financial frauds. IDPL develop Currency Transaction Report (CTR) to monitor and inform the costumer deposit fund. Furthermore, Electronic Identification, Authentication and Trust Services (EIDAS) procedures offer opportunities to combine liveness detection of the user with qualified e-signature that validate the recognition of user signature as legal equivalent of handwriting, it can be used on cheques or any banking actions that requires the signature of the clients (QTPS). IG legislation body is based on AMLD5 regulation of cryptocurrency relations and services providers. AMLD5 made compulsory crypto exchanges and e-wallet providers under the businesses or organisations financial regulations. It decreases the level of the threshold for prepaid cards and the usage of those cards for money loundering.

## 7. IDPL monitoring measures and review mechanisms

Policy 7.1 - "The company must annually examine all policy components sustainability in accordance with threats and risks."

Analysis of 7.1 - IDP applies the Principle of Last Privilege (POPL) principle, that means users must have access to the exposed minimum information needed and to accomplish their tasks. Data users should only give access only to the required files to job perform functions. Electronic Record Management (ERM) enhance the improvement of IG policy, auditing, maintenance, and testing of digital signatures and Business Process Management System. Event logs are crucial to determine the cause of the company breach. According to Smallwood, 2020

ISO 27001 highlights the necessity of System Information Event Management (SIEM) solution that collocates logs. All logs should be maintained and analysed in real time to solve the company's problems. The company uses Database Activity Monitoring (DAM) and database auditing tools to detects intrusion or suspicious activity in real time. All the database users also the high-level users as database administrator, custodian of data, developers must be monitored.

Policy 7.2 - "IDPL company should give the employee the opportunity to satisfy their tastes by allowing the usage of personal devices by applying Bring Your Own Device (BYOD) policy."

Analysis of 7.2 - Access Points (APs) have been implemented in company personnel to connect their devices via network. BYOD increases the human resources productivity, reduce employee stress, and improve company finances (Gorecki, 2020). The BYOD models followed by IDPL company are: Mobile Device Management (MDM), Mobile Application Management (MAM) and Mobile Information Management (MIM). MDM allows the company to monitor the performance and the function of mobile devices in remote manner. MDM gateway is configured on DMZ to allow mobile devices the access of corporate network services whilst the server is monitoring their performance. MIM demonstrates the limitations of installing applications in mobile devices. It comes with the integrated MAM APs solution into mobile devices. IT administrators use MAM model to install, monitor, audit and update customized applications. It allows IT administrators to take control of legitimate applications on mobile devices and to avoid the compromising of sensitive data owner or user.

## 8. Compliance

Policy 8.1 - "Staff must adhere all the policy and procedure of company in accordance with discipline regulation."

Analysis of 8.1 - It is mandatory for all the company staff members to respect all legal and ethical feature of regulations.

## 9. Approval

Policy 9.1 - "On the title page of this policy is shown the revision number and the date of approval by the bord."

Analysis of 9.1 – The title page of this paper of work includes the revision number and the date of approval.

## 5. Summary

### 5.1 The findings from previous critical analysis and recommendations

One problem faced with IDPL IG policy establishing is efficiently improved reporting and responding mechanisms need to be accomplished to improve company performance regarding to system issues. It can be developed by categorising disk space issues, update issues and backup failures. On the other hand, there are a lot of paper records that involve details from the system of clients. Those contain a high security risk if they are in wrong hands.

Some recommendation from this research related to security that should be considered as part of culture or behaviour and not as a set of regulations and rules. So, the Information Governance Policy should involve all the staff in discussions or commitment of IG issues. Also, paper documents are not necessary to be destroyed. Trivial or redundant data can be destroyed from the system so, some determined kind of emails can be deleted older than 18 months.

### 5.2 Costs and Timescale

1. The hard drive of amortized devices must be destroyed, it should be completed May 2023. The cost of one device should not be more than 2000 GBP.
2. The updating of Information System and Event Management System to collect and to detect logs activities must be done by April 2023. The cost of systems maintenance is 10000 GBP.
3. The scanning and monitoring of vulnerabilities on internal and external systems components must be setup. This is proposed to be started on 17 November 2022 and to finish on 17 November 2025 at a cost of 25000 GBP.

## 6. Conclusion

This academic research highlights the companies need to develop an Information Governance Policy in a high level of ISO/IEC 27001:2022 standard implemented within GDPR and supported by Jersey Security Strategy. IDPL agrees to JSS importance components such as a proper cultural way of staff training, risk detection, threads, and vulnerability awareness for each security management system component. IG policy helps the company to align it with regulatory bodies and developed industries security standards, to examine the current performance of the IDPL company, to get analysed feedback for the impacts of staff training commitments. The examination of IG policy offers a hight opportunity to provide optimizes solutions mechanisms for IDPL company security systems management. Once, the company ascertains suspicious incidents it feels confident to implement its framework on reasonable measures.